\documentclass[apjl]{emulateapj}
\usepackage{apjfonts}

\shortauthors{Hoard et al.}
\shorttitle{{\it Herschel} Observations of $\epsilon$~Aurigae}

\begin{document}

\title{The Invisible Monster Has Two Faces:\\ Observations of $\epsilon$~Aurigae with the {\it Herschel Space Observatory}}
 
\author{D.~W.\ Hoard\altaffilmark{1}, 
D.\ Ladjal\altaffilmark{2},
R.~E.\ Stencel\altaffilmark{2},
and S.~B.\ Howell\altaffilmark{3}} 

\altaffiltext{1}{{\it Spitzer} Science Center, California Institute of Technology, Pasadena, CA 91125}
\altaffiltext{2}{Department of Physics and Astronomy, University of Denver, Denver, CO 80208}
\altaffiltext{3}{NASA Ames Research Center, Moffett Field, CA 94035}

\slugcomment{to appear in Astrophysical Journal Letters}

\begin{abstract}
We present {\it Herschel Space Observatory} photometric observations of the unique, long-period eclipsing binary star $\epsilon$~Aurigae.  Its extended spectral energy distribution is 
consistent with our previously published cool (550~K) dust disk model.  
We also present an archival infrared spectral energy distribution of the side of the disk facing the bright F-type star in the binary, which is consistent with a warmer (1150~K) disk model.  
The lack of strong molecular emission features in the {\it Herschel} bands suggests that the disk has a low gas-to-dust ratio.
The spectral energy distribution and {\it Herschel} images imply that the 250~GHz radio detection reported by Altenhoff et al.\ is likely contaminated by infrared-bright, extended background emission associated with a nearby nebular region and should be considered an upper limit to the true flux density of $\epsilon$~Aur.  
\end{abstract}


\section{Introduction}

In \citet{paper1} (henceforth, Paper~I), we provide a summary of previous results on $\epsilon$~Aurigae, the eclipsing binary star with the longest known orbital period (27.1~yr), and introduce an empirical model for this system.
Our model was constrained by a spectral energy distribution (SED) spanning three orders of magnitude in wavelength, from the far-ultraviolet to the mid-infrared, assembled from new and archival ground- and space-based observations.  We established the nature of $\epsilon$~Aur as a large ($R\approx135$~R$_{\odot}$), but low mass ($M\approx2$~M$_{\odot}$), post-asymptotic giant branch F-type star orbiting a B5 main sequence star surrounded by a large ($R\approx4$ AU), cool (550~K) dust disk.  Subsequent to Paper~I, a number of authors have further illuminated the $\epsilon$~Aur disk (e.g., \citealt{stencel11}, \citealt{budaj11}, \citealt{chadima11}, \citealt{takeuti11}), highlighting common threads pointing to agreement with the low mass scenario and existence of discrete optically thick and thin disk components.  \citet{klop10} even obtained near-infrared interferometric images of the disk passing across the face of the F star during eclipse ingress, which have enabled better constraints on the characteristics of the optically thick disk component.

The Paper~I SED was constructed primarily from data obtained close in time to the latest (2009--2011) or previous (1982--1984) primary eclipse of $\epsilon$~Aur.  As such, it represents a view of the unilluminated ``back'' side of the disk relative to the F star.  We now present an expansion of the Paper~I SED through another order of magnitude of wavelength, via new observations with the {\it Herschel Space Observatory}\footnote{{\it Herschel} is an ESA space observatory with science instruments provided by European-led Principal Investigator consortia and important participation from NASA.} \citep{herschel}.
We also explore the properties of the ``front'' side of the disk (i.e., the side facing -- and directly illuminated by -- the F star) via archival observations obtained at orbital phases around disk opposition.

\section{Observations}

We observed $\epsilon$~Aur with the Photodetector Array Camera and Spectrometer (PACS; \citealt{pacs}) and Spectral and Photometric Imaging Receiver (SPIRE; \citealt{spire}) on {\it Herschel} (Open Time Cycle-1 program ``OT1\_hoard\_1''). These observations were accomplished on 9--10 Sep 2011 UT (PACS) and 21 Sep 2011 UT (SPIRE), shortly after the end of the optical eclipse in early July 2011 \citep{stencel11}.  
The observations consist of 2, 2, and 4 ``mini-scan maps'' of 10 repetitions in each of the PACS B (B1; 70~$\mu$m), G (B2; 110~$\mu$m), and R (160~$\mu$m) bands, respectively, with each pair at cross-scan angles of 70 and 110 degrees, and 8 ``small map'' repetitions at 250, 350, and 500~$\mu$m for SPIRE.

We first fit each image using a Gaussian profile to precisely determine the source position.  We then integrated around this position within apertures of 10, 15, 20, and 22 arcsec for the 70, 110, 160, and 250~$\mu$m images, respectively ($\epsilon$~Aur is not detected in the 350 and 500~$\mu$m images -- see below). The aperture photometry was performed on Level 2 data products (reduction pipeline v7.2.0 for PACS and v7.3.0 for SPIRE) using a custom photometry routine written by D.\ Ladjal that is based on the DAOPHOT photometry package but with error calculations adapted to bolometers.  We applied corrections for both aperture (from the online PACS and SPIRE photometry cookbooks) and color (based on the slope of the photometric points as a function of wavelength).

\begin{figure*}[tb]
\epsscale{0.97}
\plotone{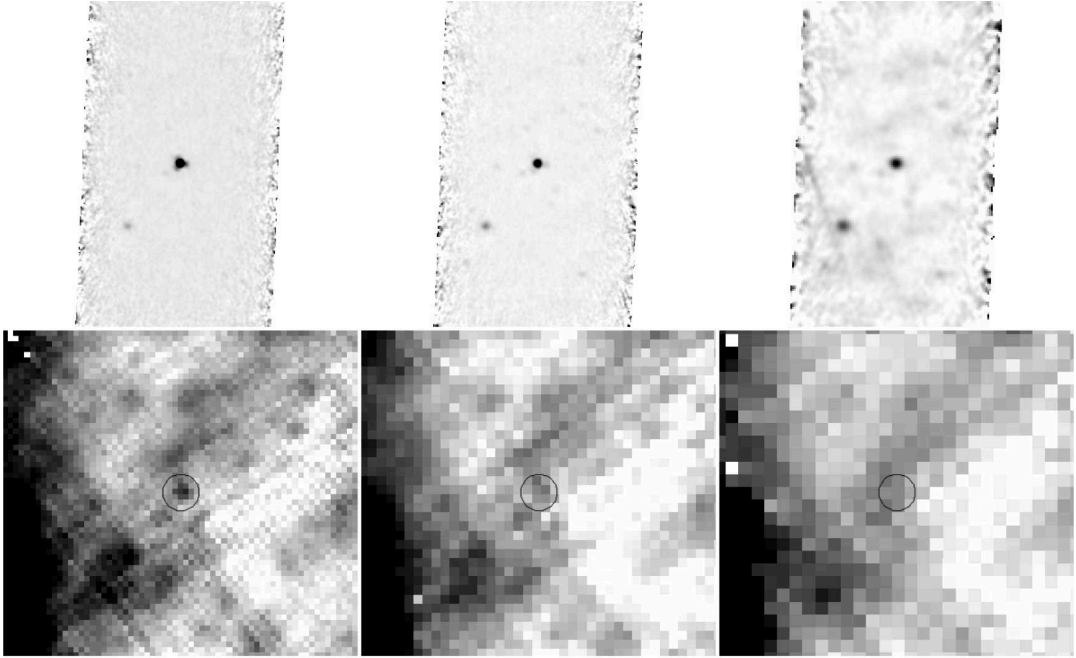}
\epsscale{1.0}
\caption{{\it Herschel} Level 2 pipeline images of $\epsilon$~Aur 
consisting of combined PACS cross-scan images at 70, 110, and 160~$\mu$m (top row) and SPIRE maps at 250, 350, and 500~$\mu$m (bottom row; wavelengths increasing from left to right in both rows).  
The SPIRE images are shown with a logarithmic greyscale stretch, while the PACS images are linear.
The SPIRE images are 6.4 arcmin wide by 5.9 arcmin tall (the circles marking the location of $\epsilon$~Aur have a diameter of 40 arcsec); the PACS images have the same angular scale as the SPIRE images.
All of the images are oriented with north up and east to the left, rectified to the world coordinate solution of the PACS 70 $\mu$m image.
\label{f:images}}
\end{figure*}

Figure \ref{f:images} shows images from each {\it Herschel} band, and the photometry is listed in Table~\ref{t:data}.  The photometric uncertainties are quadrature sums of the absolute measurement errors (0.09\%, 0.3\%, 1.8\%, 21\% for 70, 110, 160, 250 $\mu$m, respectively) and the instrumental and calibration errors (5\% for PACS-70, {\mbox -110}; 10\% for PACS-160; 15\% for SPIRE).  As a check of the {\it Herschel} photometry, we note that the PACS-70 value differs by only 1\% from the {\it Spitzer} MIPS-70 value from Paper~I (listed here in Table~\ref{t:data}), well within their mutual $1\sigma$ uncertainties (5\% and 15\%, respectively).

The upper limits in Table~\ref{t:data} at 350 and 500 $\mu$m are $3\sigma$ r.m.s. values for each map (in mJy/beam).  
Inspection of the images suggests that these non-detections can be blamed, at least in part, on the presence of infrared-bright structured background emission, rather than simply faintness of $\epsilon$~Aur at these wavelengths.  
This structured background is likely related to the known dark cloud nebular region TGU~H1105 \citep{2005PASJ...57S...1D} centered $\sim$11 arcmin northeast of $\epsilon$~Aur and extending southward -- this is visible as the very bright extended emission along the east edge of the SPIRE maps.  The point source visible in the PACS images $\approx90$ arcsec southeast of $\epsilon$~Aur (J2000 coordinates of 05:02:03.6, +43:48:15.9 measured from the PACS-160 image) is likely an embedded star or protostar, since it is not visible in optical (DSS) or near-infrared (2MASS) images\footnote{It is also not visible in the {\it Wide-Field Infrared Survey Explorer} ({\it WISE\/}) Preliminary Release Atlas images at 3.4 and 4.6~$\mu$m.  A faint source is possibly present at this position in the 12 and 22~$\mu$m {\it WISE} images, but is not listed in the photometry catalog, indicating that, if real, it has S/N$<$7 in both bands and is fainter than $\approx1$ and $\approx6$~mJy, respectively.}.

\section{Discussion}
\subsection{Orbital Geometry}

The $\epsilon$~Aur binary orbit has a large eccentricity (0.23; \citealt{stefanik10}).  Consequently, times of primary eclipse and opposition phase of the disk do not occur exactly $0.5P_{\rm orb}$ apart, as they would for a circular orbit.  Instead, \citet{stefanik10} showed that the red-to-blue crossing of the F star radial velocity curve (which is coincident with the photometric eclipse mid-point) occurs $\approx0.1P_{\rm orb}$ after the periastron passage of the disk around the F star, and the blue-to-red crossing occurs another $\approx0.6P_{\rm orb}$ later (e.g., see Figures 1 and 4 in \citealt{stefanik10}).  Thus, when using a linear orbital ephemeris with zeropoint at mid-eclipse, the disk opposition occurs at $\phi\approx0.6$.  
To determine when the data were obtained relative to the viewing geometry of the binary,
 we use a linear ephemeris constructed by combining the best orbital period found by \citet{stefanik10} with the latest eclipse midpoint date found by \citet{stencel11} (see Table~\ref{t:data}).

\begin{deluxetable*}{lllllll}
\tablewidth{0pt}
\tablenum{1}
\tablecaption{Photometry of $\epsilon$~Aur \label{t:data}} 
\tablehead{
\colhead{Wavelength} & 
\colhead{Instrument/} & 
\colhead{Observatory} & 
\colhead{Date of Observation} &
\colhead{Orbital} &
\colhead{Flux Density} &
\colhead{Reference} \\
\colhead{($\mu$m)} & 
\colhead{Band} & 
\colhead{ } & 
\colhead{(JD-2400000)} & 
\colhead{Phase\tablenotemark{a}} &
\colhead{(mJy)} &
\colhead{ }
}
\startdata
\phn\phn71.44 & MIPS-70 & {\it Spitzer}  & 53639, 53790 & $\approx$0.83 & 497$\pm$74        & Paper~I \\
   \phn\phn70 & PACS    & {\it Herschel} & 55814        &         0.043 & 492$\pm$25        & this work \\
   \phn110    & PACS    & {\it Herschel} & 55814        &         0.043 & 254$\pm$13        & this work \\
   \phn160    & PACS    & {\it Herschel} & 55814        &         0.043 & 105$\pm$25        & this work \\
   \phn250    & SPIRE   & {\it Herschel} & 55826        &         0.044 & \phn57$\pm$15        & this work \\
   \phn350    & SPIRE   & {\it Herschel} & 55826        &         0.044 & $<$60 ($3\sigma$) & this work \\
   \phn500    & SPIRE   & {\it Herschel} & 55826        &         0.044 & $<$57 ($3\sigma$) & this work \\
      1200    & radio   & IRAM 30-m      & 47710--47863 & $\approx$0.23 & \phn\phn9$\pm$2         & \citet{1994AandA...281..161A}
\enddata
\tablenotetext{a}{Calculated using the ephemeris JD$_{\rm obs}$ = JD~2455390(10) + 9896(1.6)E (mid-eclipse date from \citealt{stencel11}, orbital period from \citealt{stefanik10}).}
\end{deluxetable*}

\subsection{Cool Side of the Disk}

\begin{figure*}[tb]
\epsscale{0.90}
\plotone{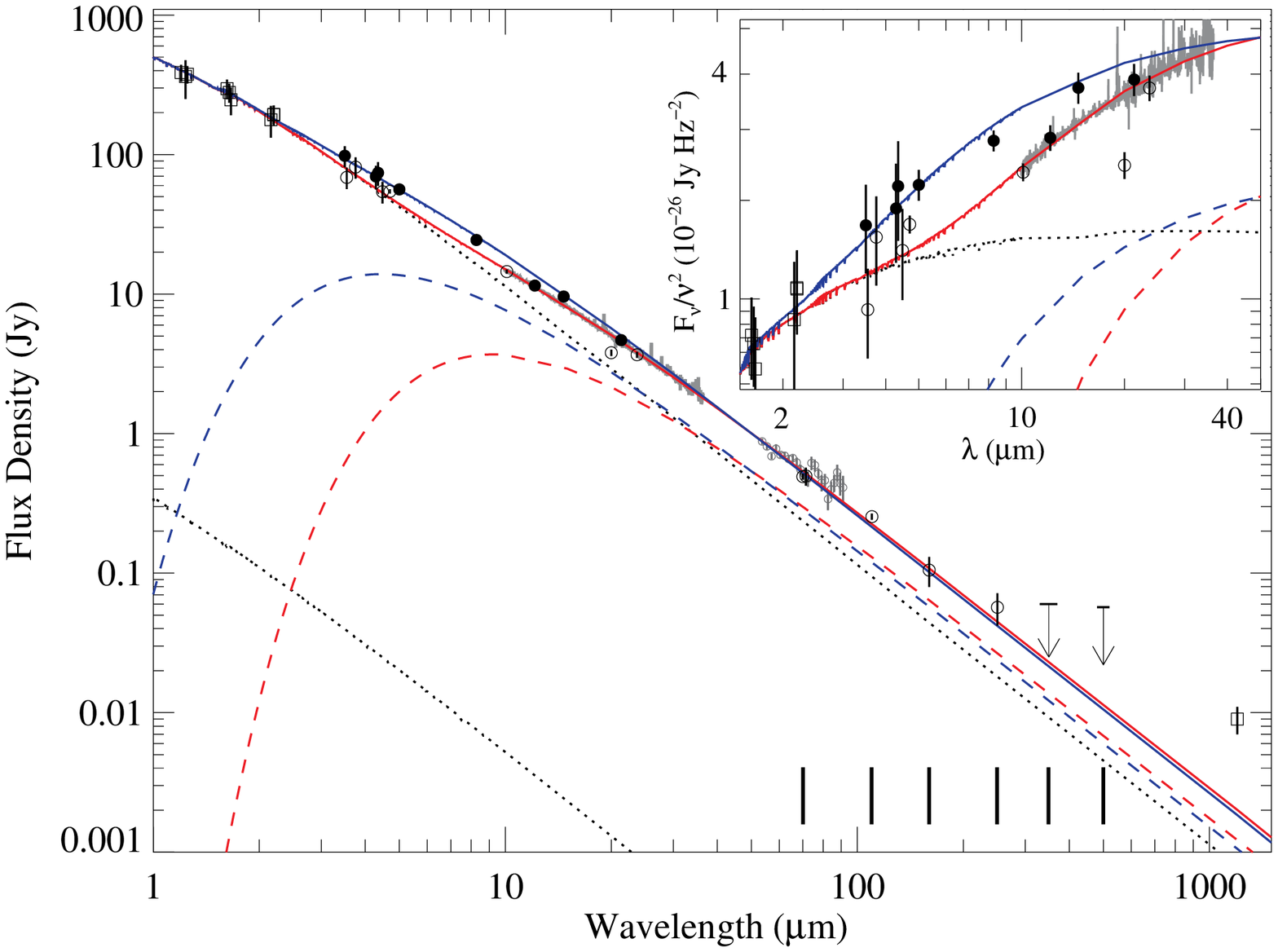}
\epsscale{1.0}
\caption{Long wavelength SED of $\epsilon$~Aur, showing measurements obtained at orbital phases corresponding to the cool (pre- and post-eclipse) and warm (opposition phase) sides of the dust disk.
The data are as follows:\ 
J, H, K/Ks from 2MASS (Paper~I), \citet{1984ApJ...284..799B} and \citet{2001AstL...27..338T}, 
as well as the 250~GHz radio detection \citep{1994AandA...281..161A} (unfilled squares).
The data corresponding to the cool disk case (unfilled circles) include:\ 
{\it Spitzer} IRAC 3.5, 4.5 $\mu$m and MIPS 24, 70 $\mu$m (Paper~I), 
ground-based L$^{\prime}$, M, N, Q \citep{1984ApJ...284..799B},
and our new {\it Herschel} PACS and SPIRE (plotted as upper limits for SPIRE 350 and 500 $\mu$m).
The {\it Spitzer} IRS and MIPS-SED spectroscopic data (Paper~I) are plotted in grey.
The data corresponding to the warm disk case (filled circles) include:\ 
ground-based L, M \citep{2001AstL...27..338T}
and the {\it MSX} B1, B2, A, C, D, E bands \citep{2003yCat.5114....0E}.
The vertical bars mark the {\it Herschel} PACS and SPIRE bands.
Two models are plotted:\ the cool case model (lower, red solid line) from Paper~I is the sum of limb-darkened model F0 (post-AGB) and B5~{\rm V} spectra (upper and lower dotted lines, respectively), and a cool (550~K) blackbody disk (lower, red dashed line); the warm case model (upper, blue solid line) is the same but utilizes a 1150~K disk (upper, blue dashed line).
The inset panel is an expanded view of the 1--50 $\mu$m region, with the y-axis scaled by $\nu^{-2}$ (i.e., divided by the slope of the Rayleigh-Jeans tail of the blackbody function) to emphasize the difference between the cool and warm cases, which are essentially indistinguishable outside of this wavelength range.
\label{f:sed}}
\end{figure*}

Figure \ref{f:sed} shows an SED of $\epsilon$~Aur derived from the one presented in Paper~I,
but starting in the near-infrared at 1~$\mu$m.  We increased the long wavelength limit compared to Paper~I by including both the new {\it Herschel} photometry and the 250~GHz radio detection reported by \citet{1994AandA...281..161A}.
We also differentiated the data listed in Table~1 of Paper~I into two groups that were measured at $\phi\approx0.85$--$0.15$ (as in Paper~I, the SED excludes data obtained during eclipse, at $\phi=$0.97--0.03; e.g., see \citealt{stencel11}) and at $\phi\approx$0.5--0.7.  The former data correspond to the cool back side of the disk (facing away from the F star) during pre- and post-eclipse, while the latter were obtained around opposition phase and correspond to the warm front side of the disk (directly facing the F star).
The back side data exclude the \citet{2001AstL...27..338T} and Midcourse Space Experiment ({\it MSX}) photometry from Paper~I (which were obtained at $\phi\approx0.6$). The exceptions to this division of the data are the near-infrared photometric points from Paper~I, all of which were obtained at orbital phases close to 0.6.  However, the disk contribution is less than a few hundredths of a percent of the total model shortward of K (less than 3\% even in the warm -- hence, bright -- disk case; see below); thus, the near-infrared photometry is effectively phase-independent.

Figure \ref{f:sed} also shows the model from Paper~I, extended to 1500~$\mu$m.  
This model utilizes a parametric description of a homogeneous, partially transparent dust disk that is characterized by a ``transmissivity'' (the degree of transparency of the disk) and an ``emissivity'' (the degree to which light radiated by dust grains inside the disk is visible from outside the disk, not to be confused with the standard emissivity $\epsilon$).  Both parameters are simple multiplicative scale factors applied to a single-temperature blackbody disk model.  
This model is effectively a bulk average over a disk composed of discrete optically thick and thin regions, that minimizes the need to constrain a larger number of parameters corresponding to multiple disk components.
The {\it Herschel} photometry is completely consistent with the 550~K dust disk model from Paper~I;
however, the radio point is $3.5\sigma$ brighter than the model.

The $\epsilon$~Aur dust disk almost certainly does contain a mix of optically thick and thin regions; \citet{stencel11} and \citet{budaj11} have discussed this, and \citet{klop10} have estimated likely geometric parameters for the optically thick component based on their interferometric imaging.  Similarly, the actual dust disk should obey a radial temperature profile characterized by higher temperature closer to the central star, even if the observational signature of this is likely to be relatively minimal, as the disk is viewed at high inclination ($i\approx89^{\circ}$; Paper~I), so its SED is dominated by the geometrically thick edge of the disk.  
As a first step toward a more ``physical'' disk model for $\epsilon$~Aur (a full treatment is beyond the scope of this paper), we modified the detailed dust disk model described in \citet{hoard09} and references therein, which was developed for modeling dust disks around white dwarfs and cataclysmic variables, to calculate a two-component (optically thick and thin) model for the $\epsilon$~Aur disk.  

Using the values derived in Paper~I and \citet{klop10}, we constrained the optically thick component to have outer radius of 3.8~AU, thickness of 0.76~AU, and outer edge temperature of 550~K, with an inclination of $89^{\circ}$ at a distance of 625~pc.  The inner edge was constrained to have a temperature of 1500~K (the assumed dust sublimation temperature).  Given the 15,000~K B5~{\rm V} central star described in Paper~I, and assuming a radial temperature profile exponent of 0.56, this results in an inner disk radius of 0.55~AU, consistent with the discussion in Paper~I concerning a possible ``hole'' in the dust disk where the central star sublimates the dust into gas.  The slight increase in the radial temperature profile exponent compared to the expected value of 0.5 (corresponding to the surface of an opaque blackbody disk), suggests that the surface of the disk is not completely opaque down to some geometric depth at which the optical depth becomes large.  This is comparable to the effect encapsulated in the ``transmissivity'' and ``emissivity'' parameters of the Paper~I model. 

The optically thin disk component is modeled as a corona or atmosphere of hotter dust grains extending 0.1~AU above and below the optically thick disk component.  This component is assumed to be composed of 50-micron silicate (3~g~cm$^{-3}$) grains, constrained to the same inner edge boundary conditions as the optically thick component, but having an outer edge temperature of 1000~K (corresponding to $R=1.9$~AU).  The total mass of dust in this component is $0.0007$~M$_{\oplus}$, 1\% of the mass estimated by \citet{klop10} for the optically thick disk component in $\epsilon$~Aur.  The combined contribution of these two model components produces a total SED almost indistinguishable from the Paper~I cool disk model, differing by, at most, 2.8\% (at 5~$\mu$m) and by less than 1\% over 85\% of the 1--1500~$\mu$m wavelength range.
Although far from a comprehensive system model for $\epsilon$~Aur (e.g., it lacks strong constraints on the additional degrees of freedom introduced by the increased physical complexity), the agreement between this calculation and the ``bulk-averaged'' model from Paper~I provides some confidence that the latter is useful for making reasonable estimates of the disk parameters in $\epsilon$~Aur.

\subsection{Warm Side of the Disk}

Using the warm side data plotted in Figure \ref{f:sed}, we calculated a new version of the Paper~I model to reproduce the SED of the front side of the $\epsilon$~Aur disk (i.e., the side closer to, and directly illuminated by, the F star).  This model is identical to the cool disk model shown here and in Paper~I, except that the disk temperature is increased to 1150~K and the ``emissivity'' factor is set to 1.00 (down from 2.43 in the cool disk model).  The disk temperature uncertainty, determined by noting the range over which the match between model and data becomes noticeably worse, is $\pm$50~K.  The 250~GHz radio point \citep{1994AandA...281..161A} is still several $\sigma$ brighter than the warm disk model, which does not differ significantly from the cool disk model at very long wavelengths.  We note in passing that the relatively faint 12 and 20 $\mu$m {\it MSX} points in the warm case data -- echoed by the low Q band point from the cool case data -- might point to the presence of silicate absorption features that were visible in the orbital cycle leading up to the 1982--1984 eclipse (when the photometric data were obtained) but not, apparently, visible in the IRS spectrum obtained prior to the latest eclipse.

The higher temperature is an expected consequence of irradiation from the F star.  The lower ``emissivity'' might be understood if increased radiation pressure from the F star on the front side of the disk compresses the disk material, effectively increasing its optical depth to the point at which the observable side of the disk at opposition is completely opaque.  This opacity change could also account for the steepness of photometric egress compared to ingress -- see the light curve in Figure~2b of \citet{stencel11}.  For completeness, we note that an alternative scenario is that the ``emissivity'' could be $>1$ if the warm disk emitting area is not completely opaque, but is smaller than the cool side emitting area.  Exploring this possibility in a more quantitative fashion is precluded by the sparseness of the data at the longest wavelengths in the warm case SED.
As described in Paper~I, the disk ``transmissivity'', which is unchanged in the warm model compared to the cool model, is primarily important in the ultraviolet and has little effect on the model's match to the infrared data.

\section{Discussion}

\citet{takeuti11} presents an analytic model for the disk in $\epsilon$~Aur that accounts for both the geometry and rotation of the disk and irradiation by the F star.  In Takeuti's model, the initial system parameters are based on our empirical disk model from Paper~I, but the disk temperature, rather than being defined, is a calculated quantity.  For a particular assumed value for the heat capacity of the disk material that produces a temperature of $\approx$500~K on the back side of the disk (comparable to the empirical result from Paper~I and this work), Takeuti's model predicts a disk front side temperature of $\approx$1200~K, in close agreement with our empirical result of $1150\pm50$~K.

In light of the {\it Herschel} extension to the $\epsilon$~Aur SED, which closely follows the disk model from Paper~I and the optically thick/thin model presented here, the \citet{1994AandA...281..161A} radio detection should be re-examined.  Admittedly, we cannot rule out the possible existence of a very cold, or non-thermal, component that contributes significantly only at wavelengths longer than $\sim1000$~$\mu$m.  In addition, because the radio observation was made at an orbital phase close to disk quadrature, it is unclear whether it is better associated with the cool or warm side of the disk.  The long wavelength end of the former is better constrained by data than that of the latter; thus, while the radio point seems incongruous in comparison to the rest of the cool disk SED, again, we cannot rule out that it originates from an unexpected component that is only visible on the warm side of the disk.  However, we suggest that a more likely scenario is that the 250~GHz radio detection (which had a half-power beam width of $\approx12$ arcsec) is contaminated by -- or potentially due solely to -- the infrared-bright structured background in the vicinity of $\epsilon$~Aur visible in the {\it Herschel} SPIRE maps (see Figure ~\ref{f:images}).  In that regard, the \citet{1994AandA...281..161A} value should be considered an extreme upper limit to the true flux density of $\epsilon$~Aur at this long wavelength.  Based on our model, we predict a much lower 1200~$\mu$m flux density of $\approx2$~mJy for $\epsilon$~Aur.

\section{Conclusions}

A conclusion that can be drawn immediately from the extended cool disk SED of $\epsilon$~Aur in Figure \ref{f:sed} is that any potentially bright emission features from molecular species in the PACS and SPIRE bands are minimal, given the adherence to the 550~K disk model. This leads us to suggest that the gas-to-dust ratio in the $\epsilon$~Aur disk is very low.  The disk itself is likely of low mass even when the gas content is included ($\ll1$~M$_{\odot}$, as suggested in Paper~I and \citealt{klop10}), and is perhaps analogous to the transitional disk in HD~95881 \citep{2010AandA...516A..48V}.

During the course of the several years between the end of primary eclipse (late 2011), disk quadrature (circa 2019), and secondary eclipse (circa 2027), the side of the $\epsilon$~Aur disk that is directly illuminated and heated by the F star is increasingly visible.  Infrared and/or sub-mm observations during this interval, providing longitude by longitude characterization of the disk's azimuthal temperature profile, can uniquely address disk thermal gradients that are not feasibly observed at any other time until after the next eclipse.  We urge observers to seize this rare opportunity.

\acknowledgements{
This work is based on observations made with {\it Herschel}, an ESA Cornerstone Mission with significant participation by NASA, and with the {\it Spitzer Space Telescope}, which is operated by the Jet Propulsion Laboratory, California Institute of Technology, under a contract with NASA.  
Support for this work was provided by NASA through an award issued by JPL/Caltech.  
We used data products from the Two Micron All Sky Survey, a joint project of the University of Massachusetts and the Infrared Processing and Analysis Center/Caltech, funded by NASA and NSF, and utilized the SIMBAD database, operated at CDS, Strasbourg, France, and NASA's Astrophysics Data System.  
We thank Sean Carey for a helpful discussion about {\it MSX} photometry.

{\it Facilities:} 
\facility{AAVSO}, 
\facility{Herschel (PACS, SPIRE)}, 
\facility{IRAS}, 
\facility{MSX}, 
\facility{Spitzer (IRAC, IRS, MIPS)}}

\end{document}